\documentclass[runningheads]{llncs}
\usepackage{float}
\usepackage{graphicx}
\usepackage{url}
\usepackage{booktabs}
\usepackage{boldline}
\usepackage{multirow}
\usepackage[ruled,vlined,linesnumbered]{algorithm2e}
\usepackage{amssymb,amsmath,bm}
\usepackage[framemethod=TikZ]{mdframed}
\mdfdefinestyle{MyFrame}{%
	linecolor=blue,
	outerlinewidth=1.0pt,
	roundcorner=20pt,
	innertopmargin=\baselineskip,
	innerbottommargin=\baselineskip,
	innerrightmargin=20pt,
	innerleftmargin=20pt,
	backgroundcolor=gray!50!white}
\usepackage{tcolorbox}
\newtheorem{mytheorem}{Theorem}
\usepackage{hyperref}
\hypersetup{
    colorlinks=true,
    linkcolor=magenta,
    filecolor=blue,      
    urlcolor=blue,
    citecolor=blue,
    pdftitle={Sharelatex Example},
    bookmarks=true,
    }
\begin{document}
\title{A Social Distancing-Based Facility Location Approach for Combating COVID-19}
\titlerunning{Social Distancing-Based Facility Location Problem}
%
\author{Suman Banerjee\inst{1} \and
Bithika Pal \inst{2} \and
Maheswar Singhamahapatra\inst{3}}
\authorrunning{S. Banerjee et al.}
%
\institute{Indian Institute of Technology, Jammu, India.
\and
Indian Institute of Technology, Kharagpur, India. \\
 \and
School of Business, FLAME University, India. \\
\email{suman.banerjee@iitjammu.ac.in, bithikapal@iitkgp.ac.in, msmahapatra@gmail.com}
}
\maketitle              
\begin{abstract}
 In this paper, we introduce and study the problem of facility location along with the notion of \emph{`social distancing'}. The input to the problem is the road network of a city where the nodes are the residential zones, edges are the road segments connecting the zones along with their respective distance. We also have the information about the population at each zone, different types of facilities to be opened and in which number, and their respective demands in each zone. The goal of the problem is to locate the facilities such that the people can be served and at the same time the total social distancing is maximized. We formally call this problem as the \textsc{Social Distancing-Based Facility Location Problem}. We mathematically quantify social distancing for a given allocation of facilities and proposed an optimization model. As the problem is \textsf{NP-Hard}, we propose a simulation-based and heuristic approach for solving this problem. A detailed analysis of both methods has been done. We perform an extensive set of experiments with synthetic datasets. From the results, we observe that the proposed heuristic approach leads to a better allocation compared to the simulation-based approach.


\keywords{Facility Location \and Social Distancing \and Integer Programming  \and COVID-19.}
\end{abstract}
\section{Introduction} \label{Sec:Intro}

In recent times, the entire world is witnessing the pandemic Corona Virus Infectious Diseases (abbreviated as COVID-19) and based on the \url{worldometers.com} data, more than two hundred countries and territories are affected \footnote{\url{https://www.worldometers.info/coronavirus/countries-where-coronavirus-has-spread/}.  Currently (around May 2021) the second wave has attacked countries like , As of now (dated 9th May, 2021), the number of deaths due to COVID-19 are $3304113$, and the  number of confirmed cases are $158863033$}. In this situation, every country tried with their best effort to combat this pandemic. `Isolation', `Quarantine', `wearing of face mask' and `Social Distancing' are the four  keywords recommended by the \emph{World Health Organization} (henceforth mentioned as WHO) to follow for combating this spreading disease. As per the topic of our study, here we only explain the concept of `Social Distancing'. As per the WHO guidelines, the Corona virus spreads with the droplets that come out during the sneezing of an already infected person \cite{tahir2019stability}. Hence, it is advisable to always maintain a distance of $1$ meter (approximately $3$ feet) from every other person. 
\par Most of the countries (including India) were under temporary lock down and all the essential facilities including shopping malls, different modes of public transport, etc. were completely closed. However, for common people to survive some essential commodities (e.g., groceries, medicines, green vegetables, milk, etc.) are essential. It is better if the state administration can facilitate these essential services among the countrymen. However, as per the WHO guidelines, it is always advisable to maintain a distance of $1$ meter from others whenever a person is out of his residence. Hence, it is an important question where to locate these facilities such that the demands of the customers can be served, and at the same time, social distancing among the people is maximized. Essentially, this is a facility location problem along with an additional measure `social distancing'. 
\par The study of the \emph{Facility Location} problem  goes back to the $17$-th century and at that time this problem was referred to as the \emph{Fermat-Weber Problem} \footnote{\url{https://en.wikipedia.org/wiki/Geometric_median}}. This problem deals with selecting the location for placement of the facilities to best meet the demanded constraints of the customers \cite{owen1998strategic}. Since its inception due to practical applications, this problem has been well studied from different points of view including algorithmic \cite{hassin2020approximability}, computational geometry \cite{bespamyatnikh2000optimal}, the uncertainty of user demands \cite{marinakis2016hybrid}, etc. However, to the best of our knowledge, the concept of social distancing is yet to be formalized mathematically, and also the facility location problem has not been studied yet with social distancing as a measure. It has been predicted that the third wave of COVID-19 is most likely to hit India sometime at the end of November or early December this year \footnote{ \url{https://www.indiatoday.in/coronavirus-outbreak/story/india-third-wave-of-covid-19-vaccine-prevention-1799504-2021-05-06} }. Hence, it is important to study the facility location problem considering social distancing as an important criterion. 
\par In this paper, we study the problem of facility location under the social distancing criteria. The input to the problem is a road network (also known as graph) of a city where the nodes are small residential zone, edges are the streets connecting the zones. Additionally, we have how many different kinds of facilities are there and in which number they will be opened, population and demand of different kinds of facilities of every zone. The goal of this problem is locate the facilities in such a way that the basic amenities can be distributed to the people and the social distancing (mathematical abstraction is deferred until Section \ref{Sec:POM}) is maximized. In particular we make the contributions in this paper.
\begin{itemize}
    \item We mathematically formalize the notion of `social distancing' and integrate it into the facility location problem. We formally call our problem as the \textsc{Social Distancing-Based Facility Location} Problem.
    \item We propose an optimization model for this problem which is a quadratic binary integer program.
    \item We propose two different approaches for solving the optimization problem. The first one is a simulation-based approach, and the other one is a heuristic solution.
    \item The proposed solution methodologies have been analyzed for time and space requirements.
    \item Finally, the proposed methodologies have been implemented with synthetic datasets and perform a comparative study between the proposed methodologies.
\end{itemize}
The remaining part of this paper is organized as follows: Section \ref{Sec:RW} presents some relevant studies from the literature. Section \ref{Sec:POM} describes our system's model, quantification of the `social distancing', and introduce the \textsc{Social Distancing-Based Facility Location Problem}. Section \ref{Sec:MMF} describes the mathematical model of our problem. In Section \ref{Sec:Solu_Appo}, we propose one simulation-based approach and a heuristic solution approach for solving this problem. Section \ref{Sec:EE} contains the experimental evaluation, and finally, Section \ref{Sec:CFD} concludes our study.


\section{Related Work} \label{Sec:RW}
The problem of facility location was initially originated from the operations research community \cite{hassin1991improved} and has been investigated by researchers from other communities as well including  computational geometry \cite{bespamyatnikh2000optimal}, graph theory \cite{tamir2001k}, Management Sciences \cite{ross1977modeling}, Algorithm Design \cite{hassin2020approximability}, Geographical Information Systems \cite{miller1996gis} and many more. Also, this problem has been studied in different settings such as Uncapasitated \cite{krishnaswamy2016inapproximability}, Capasited \cite{tran2017hypergraph}, Multistage \cite{biajoli2019biased}, Hierarchical \cite{helber2016hierarchical}, Stochastic Demand \cite{marinakis2016hybrid}, Unreliable Links \cite{narayanaswamy2018facility}, Under Disruptions \cite{afify2019evolutionary} and their different combinations \cite{ortiz2018multi}. The problem has got applications in different domains of society including health care \cite{ahmadi2017survey}, defense \cite{lessin2018bilevel}, etc. However, to the best of our knowledge, there does not exist any literature that studies the facility location problem considering the social distancing as a measure.
\par The variant of facility location introduced and studied in this paper comes under the facility location under disruption. There exists an extensive set of literature in this context. Recently, Afify et al.  \cite{afify2019evolutionary} studied the reliable facility location problem under disruption, which is the improvisation of the well-studied Reliable $p$-Median Problem \cite{mladenovic2007p} and Reliable Uncapacitated Facility Location Problem. They developed an evolutionary learning technique to near-optimally solve these problems. Yahyaei and Bozorgi-Amiri \cite{yahyaei2019robust} studied the problem of relief logistics network design under interval uncertainty and the risk of facility disruption. They developed a mixed-integer linear programming model for this problem and a robust optimization methodology to solve this. Akbari et al. \cite{akbari2017tri} presented a new tri-level facility location (also known as $r$-interdiction median model) model and proposed four hybrid meta-heuristics solution methodologies for solving the model. Rohaninejad et al. \cite{rohaninejad2018accelerated} studied a multi-echelon reliable capacitated facility location problem and proposed an accelerated \emph{Benders Decomposition} Algorithm for solving this problem. Li and Zhang \cite{li2018sample} studied the supply chain network design problem under facility disruption. To solve this problem they proposed a sample average approximation algorithm. Azizi \cite{azizi2019managing} studied the problem of managing facility disruption in a hub-and-spoke network. They formulated mixed-integer quadratic program for this problem which can be linearized without significantly increasing the number of variables. For larger problem instances, they developed three efficient particle swarm optimization-based metaheuristics which incorporate efficient solution representation, short-term memory, and special crossover operator. Recently, there are several other studies in this direction \cite{hatefi2014robust}, \cite{fotuhi2017reliable}. As mentioned previously, there is an extensive set of literature on facility location problems on different disruption, and hence, it is not possible to review all of them. Look into \cite{snyder2016or}, \cite{ivanov2017literature} for a recent survey.
\par To the best of our knowledge, the problem of facility location has not been studied under the theme of social distancing, though this is the need of the hour. Hence, in this paper, we study the facility location problem with the social distancing criteria. 
\section{Preliminaries and the System's Model} \label{Sec:POM}
In this section, we present the background and formalize the concept of social distancing. The input to our problem is the road network of a city represented by a simple, finite, and undirected graph $G(V,E,W)$. Here, the vertex set $V(G)=\{v_1, v_2, \ldots, v_n\}$ represents the small residential zones, the edge set $E(G)=\{e_1, e_2, \ldots, e_m\}$ denotes the links among the residential zones. $W$ denotes the edge weight function that maps the distance of the corresponding street, i.e., $W:E(G) \longrightarrow \mathbb{R}^{+}$. We use standard graph-theoretic notations and terminologies from \cite{west2001introduction}. For any zone $v_j \in V(G)$, $N_{< d}(v_j)$ denotes the set of neighbors of $v_j$ within the distance $d$. We denote the number of nodes and edges of $G$ by $n$ and $m$, respectively. Additionally, we have the following information. For each $v_j \in V(G)$, $P(v_j)$ denotes the people residing at zone $v_j$. In the rest of the paper, we use the term `zone' and `node', interchangeably.  Let, $\mathcal{H}=\{h_1, h_2, \ldots, h_k\}$ denotes the $k$ different kinds of  facilities (say, $h_1 \equiv \text{ `groceries' }$, $h_2 \equiv \text{ `medicine' }$ and so on) need to be opened. For every zone $v_j \in V(G)$ and every kind of facility $h_i \in \mathcal{H}$, we know the corresponding demand $\mathcal{D}_{v_j}(h_i)$. Hence, the total demand for the facility $h_i$ is $\mathcal{D}(h_i)=\underset{v_j \in V(G)}{\sum} \mathcal{D}_{v_j}(h_i)$.

\paragraph{\textbf{Social Distancing}} Now, we formally explain the meaning of the term `social distancing' and express it as a mathematical expression. Suppose, a facility $h_i \in \mathcal{H}$ is opened for a particular duration $[t_1,t_2]$ in a day at the zone $v_j \in V(G)$.  Let, $k^{t}_{i,j}$ denotes the number of people in the queue before this facility at time $t \in [t_1,t_2]$. From the geographical location, it is possible to determine the maximum number of people that can be present in the queue by maintaining WHO recommended distance to each other. For the zone $v_j \in V(G)$ and facility $h_i \in \mathcal{H}$, this number is denoted as $\gamma_{i,j}$. If the number of people in the queue is within the threshold then the social distancing should be maximum and when it crosses the threshold the social distancing factor decreases linearly or exponentially depending upon the number of people. We denote the social distancing function by $S()$. The following equation mentions this.  

\[
S(k^{t}_{i,j}) =
\begin{cases}
\text{positive constant,} & \text{if } k^{t}_{i,j} < \gamma_{i,j} \\
\text{decreases linearly or exponentially,} & \gamma_{i,j} \leq k^{t}_{i,j} < 2 \cdot \gamma_{i,j} \\
\text{decreases linearly or exponentially,} &  k^{t}_{i,j} \geq 2 \cdot \gamma_{i,j}
\end{cases}
\]

Once $k^{t}_{ij} > \gamma_{i,j}$, the value of social distancing function decreases in two ways: linearly and exponentially, which is described below.

\begin{enumerate}
    \item \textbf{Linear Social Distance Factor}: Here, $S(.)$ decreases linearly with the increasing value current number of people in the queue  at a particular facility of certain zone. We model this using Equation \ref{Eq:Linear}.
   \begin{equation} \label{Eq:Linear}
 S(k^{t}_{i,j}) = A - b*max\{k^{t}_{i,j} - \gamma_{i,j}, 0\}, \ \text{ where } A \text{ and } b \text{ are positive constant}. 
  \end{equation}

    \item \textbf{Exponential Social Distance Factor}: Here, the $S(.)$ decreases exponentially with the linear increment of the current queue strength at a particular facility of certain zone. We model this using Equation \ref{Eq:Exponential}.
  \begin{equation} \label{Eq:Exponential}
 S(k^{t}_{i,j}) = A - b \cdot [\gamma_{i,j} + exp (k^{t}_{i,j}-2 \cdot \gamma_{i,j})^{\frac{1}{4}}], \ \text{ where } A > 1 
  \end{equation}
\end{enumerate}

The goal of the social distancing-based facility location problem is to assign facilities in the zones such that the total value of the social distancing function is maximized. Formally, we present this problem as follows.
\begin{center}
\begin{tcolorbox}[title=\textsc{Social Distancing-Based Facility Location Problem}, width=12cm]
\textbf{Input:} The road network $G(V,E,W)$, Population at each zone $v_j \in V(G)$; i.e.; $P(v_i)$, Types of facilities $\mathcal{H}=\{h_1, h_2, \ldots, h_k\}$, Number of facilities of each type to be open; i.e.; $\{ \ell_{1}, \ell_{2}, \ldots, \ell_{k}\}$, and their corresponding demand in each zone $v_j$; i.e.; $\{ \mathcal{D}_{v_j}(h_1), \mathcal{D}_{v_j}(h_2), \ldots, \mathcal{D}_{v_j}(h_k) \}$ for all $v_j \in V(G)$.
\vspace{0.2 cm}
\\
\textbf{Problem:} Place maximum number of facilities to be opened of each type such that total social distancing is maximized.
\end{tcolorbox}
\end{center}
Here, we note that each $\ell_{i} << n$. Symbols and notations used in this study are given in Table \ref{tab:Symbols}.

\begin{table}
\centering
  \caption{Symbols with their interpretation}
  \label{tab:Symbols}
  \begin{tabular}{||c|c||}
    \toprule
    Symbol & Interpretation \\
    \midrule
    $G(V,E,W)$ & The road network under consideration \\
    $V(G), E(G)$ & Residential zones and road segments joining them \\
    $W$ & Edge weight function\\
    $N_{<d}(v_j)$ & Neighbors of the zone $v_j$ within distance $d$ \\
    $d_{max}$ & Maximum  degree of a node in $G$\\
   $P(v_j)$ & Population at the zone $v_j$\\
   $P_{max}$ & Maximum population of a zone in $V(G)$\\
   $P$ & Total population of all zones of $V(G)$\\
   $\mathcal{H}$ & Set of different kinds of facilities\\
   $k$ & Number of different kinds of facilities; i.e.; $|\mathcal{H}|=k$\\
   $\ell_{i}$ & Number of facilities opened of type $h_i$\\
   $\ell_{max}$ & Maximum number of facilities among all types\\
   $[t_1,t_2]$ & Operation time window of the facilities \\
   $\Delta$ & The difference between $t_2$ and $t_1$ \\
   $k^{t}_{i,j}$ & Number of people in the queue at $v_j \in V(G)$, $h_i \in \mathcal{H}$, and $t \in [t_1,t_2]$ \\
   $\gamma_{i,j}$ & Queue strength of the facility $h_i \in \mathcal{H}$ at $v_j \in V(G)$\\
   $S(.)$ & The social distancing function \\
   $\mathcal{D}_{v_j}(h_i)$ & Demand of facility type $h_i$ at zone $v_j$ \\
   \bottomrule
\end{tabular}
\end{table}

\section{Mathematical Model Formulation} \label{Sec:MMF}
Now, we formulate a mathematical model of our problem. First, we define the decision variables involved in our formulation.

\[
    x_{i,j}= 
\begin{cases}
    1,& \text{if the facility of type } h_i \text{ is opened } \text{ at zone } v_j\\
    0,              & \text{otherwise}
\end{cases}
\]

\[
    y^{t}_{i,j,p}= 
\begin{cases}
    1,& \text{if the person } p \text{ visits the facility of type } h_i \text{ at zone } v_j \text{ at time } t \\
    0,              & \text{otherwise}
\end{cases}
\]
 Given the decision variables, total social distancing can be given by the following equation. 
 \begin{equation} \label{EQ:SOCIAL_DISTANCING}
 F= \underset{h_i \in \mathcal{H}}{\sum} \ \underset{v_j \in V(G)}{\sum} \   x_{i,j} \large{(} \ \underset{t \in [t_1, t_2]}{\sum} \ \underset{p \in P(v_j)}{\sum} S(k^{t}_{i,j}) \ y^{t}_{i,j,p}  \large{)}
\end{equation}  
In Equation \ref{EQ:SOCIAL_DISTANCING}, four summations are involved. The first one is to sum up for all the facilities. Similarly, the second one is to sum up for all the nodes. Next, one is used to sum up for all distinct time stamps within the operated time interval. Finally, the last one is for all the people of a node. The goal is to maximize the function mentioned in Equation \ref{EQ:SOCIAL_DISTANCING} subject to certain constraints. Now, we present the mathematical model, and next, we describe the meaning of each constraint.
\begin{mdframed}[frametitle={Maximizing Social Distancing}, style=MyFrame]
	\begin{center}
		$max \   \underset{h_i \in \mathcal{H}}{\sum} \ \underset{v_j \in V(G)}{\sum} \   x_{i,j} \large{(} \ \underset{t \in [t_1, t_2]}{\sum} \ \underset{p \in P(v_j)}{\sum} S(k^{t}_{i,j}) \ y^{t}_{i,j,p}  \large{)} $
	\end{center}
	subject to, 
	\vspace{-1 cm}
	\begin{center}
		\begin{equation} \label{cons:1}
		\underset{v_j \in V(G)}{\sum} x_{i,j} \leq \ell_{i}, \ \ \forall h_i \in \mathcal{H}
		\end{equation}
		\begin{equation}  \label{cons:2}
		\sum  {\bm{y}^{:}_{i,j,p}}^{T}  \ \bm{Y}^{:}_{i,:,p} = 1,  \ \ \forall p \in \underset{v_{k} \in V(G)}{\cup} P(v_{k}), \ \ \forall v_j \in V(G), \ \ \forall h_i \in \mathcal{H}
		\end{equation}
		\begin{equation} \label{cons:3}
		\underset{t \in [t_1, t_2]}{\sum} \ \underset{v_k \in \{v_j\} \cup  N_{<d}(v_j)}{\sum} y^{t}_{i,k,p} \geq 1,  \forall h_i \in \mathcal{H}, \forall v_j \in V(G), \forall p \in P(v_j)
		\end{equation}
		\begin{equation} \label{cons:4}
		\begin{split}
		y^{t}_{i,j,p}	(\underset{t^{'} \in [t_1, t_2]}{\sum}  y^{t^{'}}_{i,k,p}) \{ t - \underset{t^{'} \in [t_1, t_2]}{\sum}  y^{t^{'}}_{i,k,p} {t^{'}}  \}  \le 0, & \ \forall h_i \in \mathcal{H}, \ \forall (j,k) \in E(G), \\
		& \forall p \in P(v_j), \ \forall t \in [t_1, t_2]	
		\end{split}
		\end{equation}	 
		\begin{equation} \label{cons:5}
		x_{i,j} \in \{0,1\}, \ \ \forall h_i \in \mathcal{H}, \ \ \forall v_j \in V(G)
		\end{equation}
		\begin{equation} \label{cons:6}
		y^{t}_{i,j,p} \in \{0,1\}, \ \ \forall h_i \in \mathcal{H}, \ \ \forall v_j \in V(G), \ \ \forall t \in [t_1, t_2], \ \ \forall p \in \underset{v_{k} \in V(G)}{\cup} P(v_{k})
		\end{equation}
	\end{center}
\end{mdframed}
The first constraint in Inequation \ref{cons:1} enforces that the total number of open facilities of a particular type $h_i$ should not be more than its allowance (which is $\ell_{i}$). All the remaining constraints are on $y^{t}_{i,j,p}$. The assumption is a person can visit a facility location at once. Then, he or she can wait there for a certain period for being served or visit the nearest neighboring facility location to get the service. Now, the quantification of the continuous time domain is done in such a way that $y^{t}_{i,j,p}$ can be one, only once for the entire duration of $[t_1, t_2]$. Another physical condition implies that one person can not be in two different facility locations at the same time. Both the scenarios are capture in the second constraint described in Inequation \ref{cons:2}. Here, $\bm{Y}^{:}_{i,:,p}$ is the matrix of shape total duration by the number of facility locations denoting the presence of a person $p$ at any location in the entire time duration to avail $h_i$-type facility, and $\bm{y}^{:}_{i,j,p}$ is the one hot vector of the matrix $\bm{Y}^{:}_{i,:,p}$, which describes the presence at the particular location $v_j$. Now, the next constraint in Inequation \ref{cons:3} implies that the people should be served either at the location where they stay (e.g. $v_j$), or any nearby locations (e.g. $N_{<d}(v_j)$) within certain vicinity. The next constraint in Equation \ref{cons:4} tells that if an arbitrary person $p$ visits the locations at $v_j$ and $v_{k} \in N_{< d}(v_j)$ to avail the facility type $h_i$ then he first visits the place where he stays, then the neighboring ones. Finally, Inequation \ref{cons:5} and \ref{cons:6} tells that the decision variables $x_{ij}$ and $y_{ijp}^{t}$, for all $h_i \in \mathcal{H}$, $v_j \in V(G)$, $t \in [t_1, t_2]$ take binary values.
\vspace{0.2 cm}
\\
\textbf{\underline{Note}}: Observe that in this study we do not consider the cost of the facilities which is quite natural in case of traditional facility location problems \cite{owen1998strategic}. This is due to the following reason. During the lock down period, a significant population are jobless, and hence government is providing the basic amenities free of cost. This is our assumption as mentioned in Section \ref{Sec:Intro}. We also assume that there is no dearth of supply.
\section{Proposed Solution Approaches} \label{Sec:Solu_Appo}
As the facility location problem is \textsf{NP-Hard}, finding the optimal solution for our problem is also computationally intractable. Hence, we propose two different approaches to solve our problem. The first one is the simulation-based approach and the second one is a heuristic algorithm. It is important to note that none of these methods produces an optimal solution.
\subsection{Simulation-Based Approach}  \label{Sec:Simulation}
A simulation-based approach has been adopted to obtain a solution for the developed model. Demand is generated for the people from all the nodes. As mentioned previously, a facility of type $h_{i}$ can have at most $\ell_{i}$ many ($\ell_i < < n$). People of a node get the service of a facility  from the same node if it is available there, otherwise move to the neighbor  node where a service facility is opened. People’s arrival to the service facility follows \emph{Poisson Distribution} and service time follows \emph{Exponential Distribution} \cite{asmussen2008applied}. People get service based on the first come first serve basis. We have simulated these features of how people are arriving and getting the service and during this process how social distance is being maintained based on the number of people in the queue at a given time. People from several nodes may come and form a queue based on the arrival time and get the service based on first come first serve (FCFS) basis. Due to the space limitations we do not discuss anything related to the queuing theory and can be found at \cite{asmussen2008applied}.

\begin{algorithm}[h]
\SetAlgoLined
\KwData{Underlying Road Network $G(V, E, W)$, Population and Respective Demand of Every Zone, Number of Different Types of Facilities, Number of facilities of each type}
\KwResult{Location of Different Facilities}
 $Simulation\_Run \longleftarrow 0$\;
 \While{$Simulation\_Run$ \text{ not reached }}{
  \textbf{Step 1}: For every type of facility (say $h_i$) the allowed number of them ( say $\ell_{i}$)are opened randomly\;
  \textbf{Step 2}: Arrival time of peoples of all the nodes are generated based on the predefined arrival rate.\;
   \textbf{Step 3}: For each node, where facility is not opened, the nearest open facility is found based on the least distance.\;
   \textbf{Step 4}: For a node $v_j$, the people of of this zone and those who are assigned to this facility, their arrival times are sorted and queue is formed.  
   
   \textbf{Step 5}: Service times are generated for the nodes where a given type of service facility is opened using the exponential distribution of service time.\;
   \textbf{Step 6}: Based on the arrival time and service time, number of people in queue is derived. \;
  \textbf{Step 7}: Social distance is calculated at each queue and summed up.\;
  \textbf{Step 8}: $Simulation\_Run$ is incremented by $1$; i.e.; $Simulation\_Run=Simulation\_Run +1$ \;
 }
 \textbf{Step 9}: Among all the simulation runs corresponding locations of the maximum is returned for locating the facilities.\;
 \caption{Pseudo Code for the Simulation-Based Approach}
 \label{Algo:1}
\end{algorithm}

\par Now, the arrival time of people of all the nodes is generated based on the predefined arrival rate. It is implemented based on the inter-arrival time using poisson distribution which is reciprocal of arrival rate. Next, for each node, where the facility is not opened, the nearby open facility is found based on the least distance. Arrival time for all the people in a given and its nearby nodes are sorted. Service times are generated for the nodes where a given type of service facility is opened using the exponential distribution. Social distance is calculated at each open node using Equation \ref{EQ:SOCIAL_DISTANCING}. This process is repeated for a defined number of simulation runs. Finally, the locations corresponding to the maximum value of the social distancing function are returned as a solution. This process is shown in the form of pseudo code in Algorithm \ref{Algo:1}.

\par Now, we analyze Algorithm \ref{Algo:1} to understand its time and space requirement. Let, $\mathcal{R}$ denotes the number of simulation runs. Let, the number of different facilities are $k$ and $\ell_{max}$ denotes the maximum number of facilities among all types. Now, it is easy to observe that the running time of Step 1 is of $\mathcal{O}(k \cdot \ell_{max})$. Let, $P(v_i)$ denotes the number of people of at node $v_i$ and $P$ denotes the number of people in all the nodes in $G$; i.e.; $P=\underset{v_i \in V(G)}{\sum} P(v_i)$. As, for a people generating the arrival time requires $\mathcal{O}(1)$ time, hence, time requirement of Step $2$ requires $\mathcal{O}(P)$ time. The number of nodes where the facility is not opened will be greater than equals to $(n-\ell_{max})$ and in the worst case it will be of $\mathcal{O}(n)$. Let $d_{max}$ denotes the maximum possible degree of a node in $G$; $d_{max} = \underset{v \in V(G)}{argmax} \ deg(v)$. For each node without facility to choose the nearest facility requires $\mathcal{O}(d_{max})$. So, for all the nodes and all types of facilities required computational time is of $\mathcal{O}(n \cdot k \cdot d_{max})$. This implies that the execution of Step $3$ requires $\mathcal{O}(n \cdot k \cdot d_{max})$ time. Let, $P_{max}$ denotes the maximum number of people residing at any zone, hence, $P_{max}=\underset{v \in V(G)}{argmax} \ P(v)$. As, in any facility the people of its neighbors may come, and hence, the total number of people in any facility is of $\mathcal{O}((d_{max}+1)P_{max})$. Sorting this people based on the arrival time will require $\mathcal{O}((d_{max} \cdot P_{max} +P_{max}) \log (d_{max} \cdot P_{max} +P_{max}))$ time. This implies that the execution of Step $4$ requires $\mathcal{O}((d_{max} \cdot P_{max} +P_{max}) \log (d_{max} \cdot P_{max} +P_{max}))$ time. Assuming that the generating service time requires $\mathcal{O}(1)$, executing Step $5$ requires $\mathcal{O}(k \cdot \ell_{max})$ time. For a single queue, calculating the number of people in it requires $\mathcal{O}(1)$ time. Hence, executing Step $6$ requires $\mathcal{O}(k \cdot (d_{max}+1)\cdot P_{max})$.

Assume that $t_2-t_1=\Delta$. Hence, from the objective function it is easy to follow that computing the social distancing requires $\mathcal{O}(n \cdot k \cdot \Delta \cdot P_{max})$ time. If in each iteration of the \texttt{while} loop, we update the maximum value of the social distancing and the corresponding location of the facilities of different types then after the last iteration we obtain the solution of the proposed simulation approach. This requires $\mathcal{O}(k \cdot \ell_{max})$ time. So, the total time required by the simulation procedure is of $\mathcal{O}(\mathcal{R}(k \cdot \ell_{max} + P + n \cdot k \cdot d_{max} + (d_{max} \cdot P_{max} +P_{max}) \log (d_{max} \cdot P_{max} +P_{max})+ k \cdot \ell_{max}+ k \cdot (d_{max}+1)\cdot P_{max} + n \cdot k \cdot \Delta \cdot P_{max} + k \cdot \ell_{max}))$. This reduces to $\mathcal{O}(\mathcal{R}(k \cdot \ell_{max} + P + n \cdot k \cdot d_{max} + (d_{max} \cdot P_{max} +P_{max}) \log (d_{max} \cdot P_{max} +P_{max}) + k \cdot d_{max} \cdot P_{max} + n \cdot k \cdot \Delta \cdot P_{max}))$. Additional space required by the proposed simulation approach is to store the arrival time of the people which is of $\mathcal{O}(n \cdot P_{max})$, service time of the queues which is of $\mathcal{O}(k \cdot \ell_{max})$, location the facilities which is of $\mathcal{O}(k \cdot \ell_{max})$, social distancing function value at each queue which is of $\mathcal{O}(k \cdot \ell_{max})$. So, the total space requirement is of $\mathcal{O} (n \cdot P_{max} + k \cdot \ell_{max})$. Hence, Theorem \ref{Th:1} holds.
 
\begin{mytheorem} \label{Th:1}
The running time and space requirement of the proposed simulation-based approach is of $\mathcal{O}(\mathcal{R}(k \cdot \ell_{max} + P + n \cdot k \cdot d_{max} + (d_{max} \cdot P_{max} +P_{max}) \log (d_{max} \cdot P_{max} +P_{max}) + k \cdot d_{max} \cdot P_{max} + n \cdot k \cdot \Delta \cdot P_{max}))$ and $\mathcal{O} (n \cdot P_{max} + k \cdot \ell_{max})$, respectively.
\end{mytheorem} 
\subsection{Heuristic Solution}
In this section, we propose a heuristic solution to solve our problem. First, we describe the working principle, and subsequently, we present it as a step-by-step procedure. As a first step, we sort all the nodes based on demand in descending order. Based on this heuristic for any type of facility $h_i \in \mathcal{H}$, $\ell_{i}$ many facilities will be placed  at the top demand zones. Now, it allocates the people to the facility in the following way: For the high demand locations the people of that zones are allocated to the same zone, and subsequently, we allocate the people of other zone in the reverse order such that the entire population of $V(G)$ is distributed as uniformly as possible.. Without loss of generality, assume that after sorting the nodes based on the demand, the order of the nodes are $\rho=<v_1, v_2, \ldots, v_n>$. For any zone $v_j \in \{v_1, v_2, \ldots, v_i\}$, the people of zone $v_i$ are allocated to the facility at the same zone. Now, $v_{i+1}$ is allocated to $v_i$, $v_{i+2}$ is allocated to $v_{i-1}$, and $v_{2i}$ to $v_{1}$. Similarly, $v_{2i+1}$ is allocated to $v_2$, $v_{2i+2}$ is allocated to $v_3$ and so on. Next from Step $2$ to  Step $7$ of Algorithm \ref{Algo:1} is executed. Pseudo code is given in Algorithm \ref{Algo:2}.

\begin{algorithm}[h]
\SetAlgoLined
\KwData{Underlying Road Network $\mathcal{G}(\mathcal{V}, \mathcal{E})$, Population and Respective Demand of Every Zone, Number of Different Types of Facilities, Number of facilities of each type}
\KwResult{Location of Different Facilities}
 
  \textbf{Step 1}: All the nodes of the network are sorted based on the demand of the nodes\;
  \textbf{Step 2}: People of nodes are allocated in the facilities as described.\;
  \textbf{Step 3}:Execute Step $2$ to Step $7$ of Algorithm \ref{Algo:1}.\;

 \caption{Pseudo code for the Heuristic Algorithm}
 \label{Algo:2}
\end{algorithm}

Now, we analyze Algorithm \ref{Algo:2} to understand its time and space requirement. Sorting of the nodes based on the demand requires $\mathcal{O}(n \log n)$ time. It is easy to observe that the allocation of the facilities requires $\mathcal{O}(n)$ time. As described in Section \ref{Sec:Simulation}, the total execution time from Steps $2$ to $7$ is of $\mathcal{O}(P+ n \cdot k \cdot d_{max} + (d_{max} \cdot P_{max} +P_{max}) \log (d_{max} \cdot P_{max} +P_{max})+ k \cdot \ell_{max} + k \cdot d_{max} \cdot P_{max} + n \cdot k \cdot \Delta \cdot P_{max})$ time. Hence, total time requirement of Algorithm \ref{Algo:2} is of $\mathcal{O}(n \log n + P + n \cdot k \cdot d_{max} + (d_{max} \cdot P_{max} +P_{max}) \log (d_{max} \cdot P_{max} +P_{max})+ k \cdot \ell_{max} + k \cdot d_{max} \cdot P_{max} + n \cdot k \cdot \Delta \cdot P_{max})$. It is easy to observe that the space requirement of Algorithm \ref{Algo:2} is same as Algorithm \ref{Algo:1} which is of $\mathcal{O} (n \cdot P_{max} + k \cdot \ell_{max})$. Hence, Theorem \ref{Th:2} holds. 

\begin{mytheorem} \label{Th:2}
The running time and space requirement of the proposed heuristic approach is of $\mathcal{O}(n \log n + P + n \cdot k \cdot d_{max} + (d_{max} \cdot P_{max} +P_{max}) \log (d_{max} \cdot P_{max} +P_{max})+ k \cdot \ell_{max} + k \cdot d_{max} \cdot P_{max} + n \cdot k \cdot \Delta \cdot P_{max})$ and $\mathcal{O} (n \cdot P_{max} + k \cdot \ell_{max})$, respectively.
\end{mytheorem} 

\section{Experimental Evaluation} \label{Sec:EE}
In this section, we describe the experimental evaluation of the proposed methodology. Initially, we start by describing the datasets.
\subsection{Dataset Description} 
\begin{figure}
\centering
\includegraphics[scale=0.8]{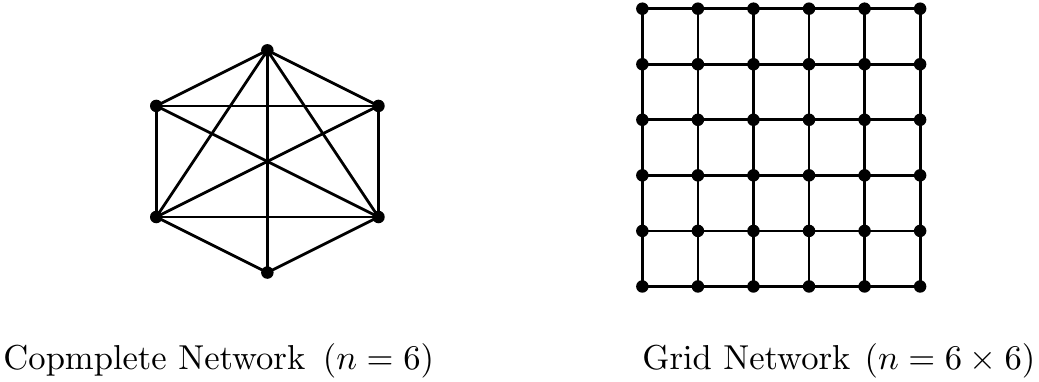}
\caption{Figure of a Complete and Grid Network of size $n=6$ and $n=6 \times 6$, respectively.}
\label{Fig:Dataset}
\end{figure}

As it is difficult to find datasets that contains all the required information in this study, hence we create some synthetic datasets. We perform our experiments with two different kinds of networks topology:
\begin{itemize}
\item \textbf{Complete Network}: In this kind of network, every vertex is connected to every other vertex of the network. It is easy to observe that an undirected complete network with $n$ vertices will have $\frac{n(n-1)}{2}$ edges.
\item \textbf{Rectangular Grid Network}: A rectangular grid network $G(V,E)$ of size $n \times n$ is defined with the vertex set  $V(G)=[n] \times [n]=\{1, 2, \ldots, n\} \times \{1, 2, \ldots, n\}$  and two vertices $(i,j)$ and $(i^{'},j^{'})$ will be connected by an edge if both $|i-i^{'}| \leq 1$ and $|j-j^{'}| \leq 1$.  Many real-world road networks are of grid structure \cite{miyagawa2009optimal,watanabe2010study}.
\end{itemize}
For the ease of understanding, we show a demo diagram of complete and grid network in Figure \ref{Fig:Dataset}.
\subsection{Experimental Setup}
In this section, we describe the experimental setup. Several parameters are there, whose value needs to be set. Here, we describe them one by one.
\begin{itemize}
    \item \textbf{Network Size}: As mentioned, we use networks of two different kinds, namely complete and grid. For complete network, we use $n=60$ and $n=100$. On the other hand for the grid network we consider $n=60 \times 60$ and $n=100 \times 100$.  
    \item \textbf{Population at Each Zone}: This has been chosen uniformly at random from the interval $[1000,2000]$ for each zone. 
    \item \textbf{Demand}: For simplicity, we consider the uniform and demand for all the people.  
    \item \textbf{Parameters Related to Queuing Theory}: We consider the inter arrival time and service time are $1$ min and $0.7$ min, respectively. 
    \item \textbf{Parameters in the Social Distancing Function}: As mentioned previously, there are three parameters in the social distancing function. They are $\gamma_{i,j}, A, \text{ and } b$. In this study, we consider $\gamma_{i,j}=4$ for all $h_i \in \mathcal{H}$ and $v_j \in V(G)$, $A=10$, and $b=0.5$.
\end{itemize}
We implement our both solution approaches in MATLAB (Version MATLAB 9.0 R2016a) and all the simulation codes and the synthetic datasets can be found at \url{https://github.com/BITHIKA1992/Facility_location_Covid19}.
\subsection{Experimental Results}
Now, we discuss the experimental results. Our goal here is to study the impact of number of facilities on social distancing and average queue length.

\begin{figure}[h]
\centering 
\begin{tabular}{ c c }
\includegraphics[scale=0.4]{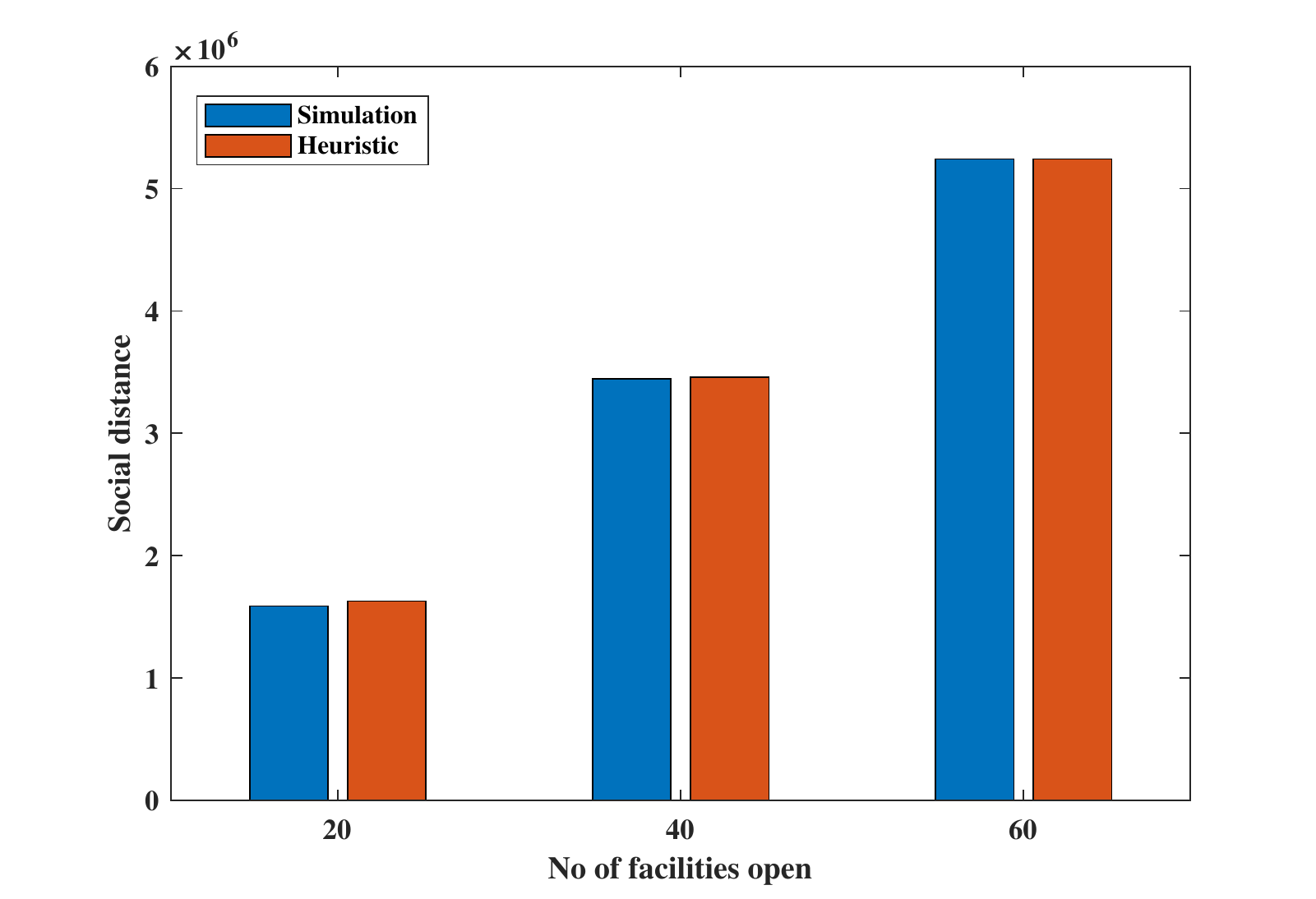} & \includegraphics[scale=0.34]{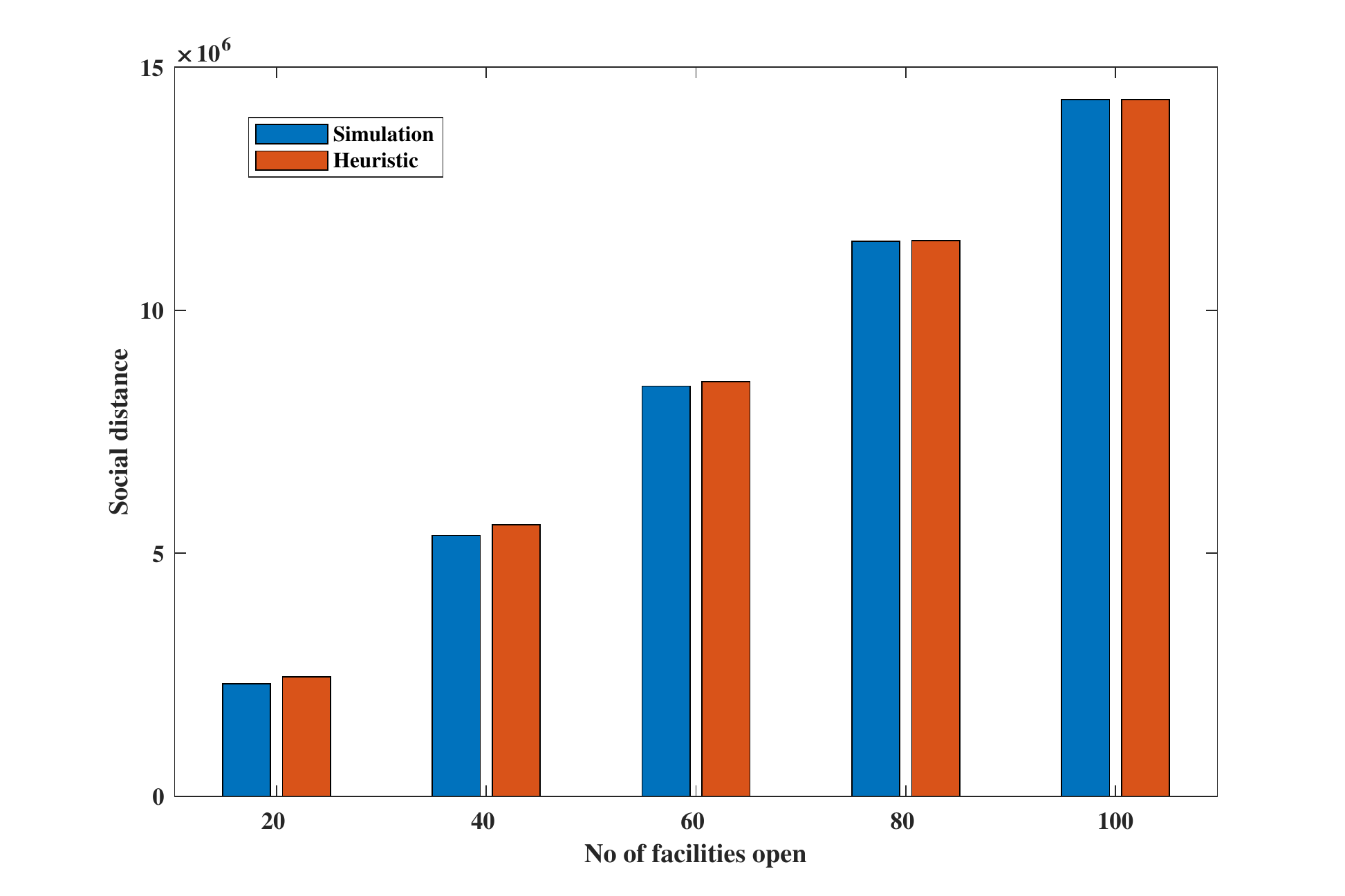} \\
(a) Complete Network ($n=60$ ) & (b) Complete Network ($n=100$)\\
\includegraphics[scale=0.4]{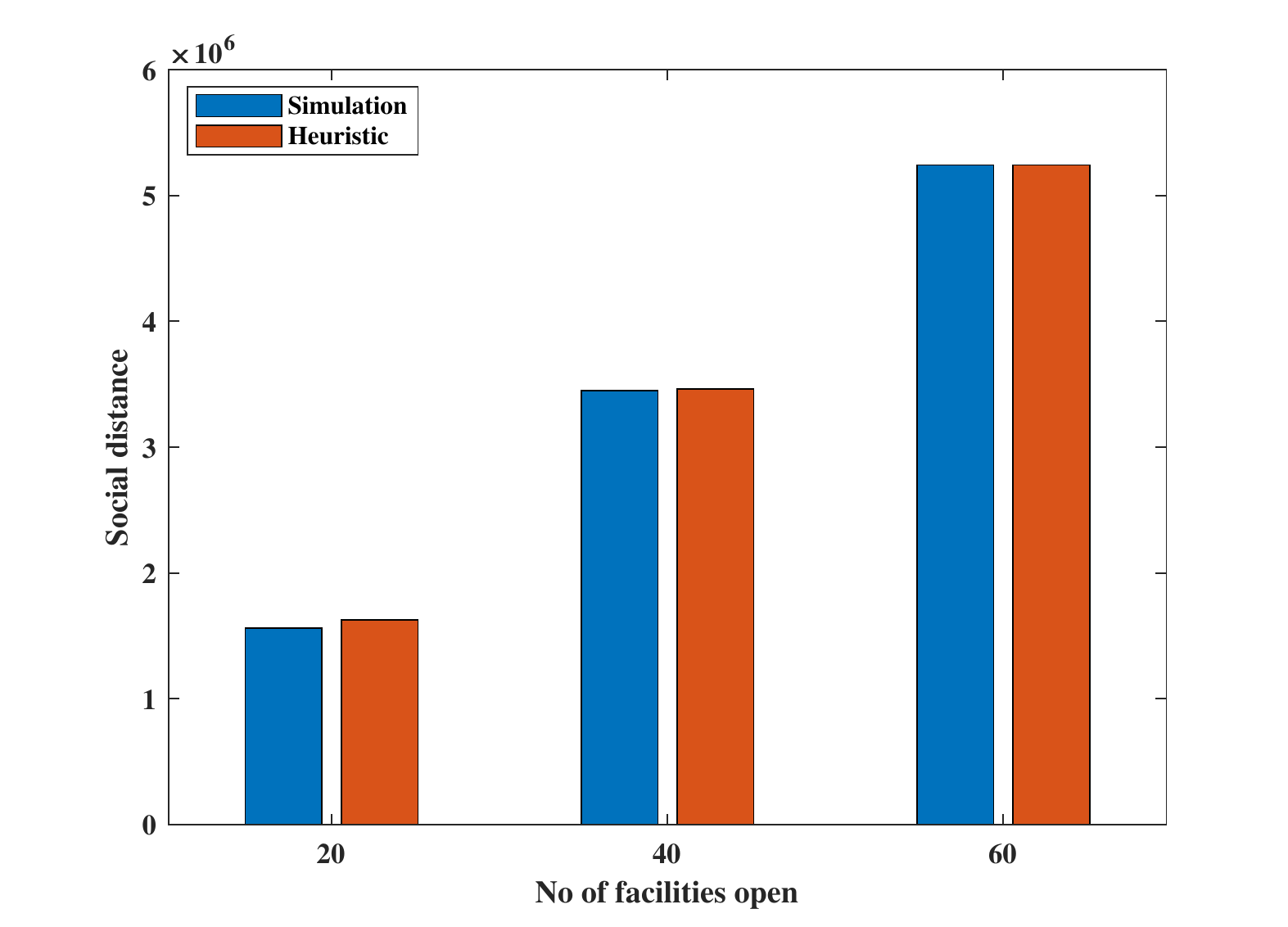} & \includegraphics[scale=0.4]{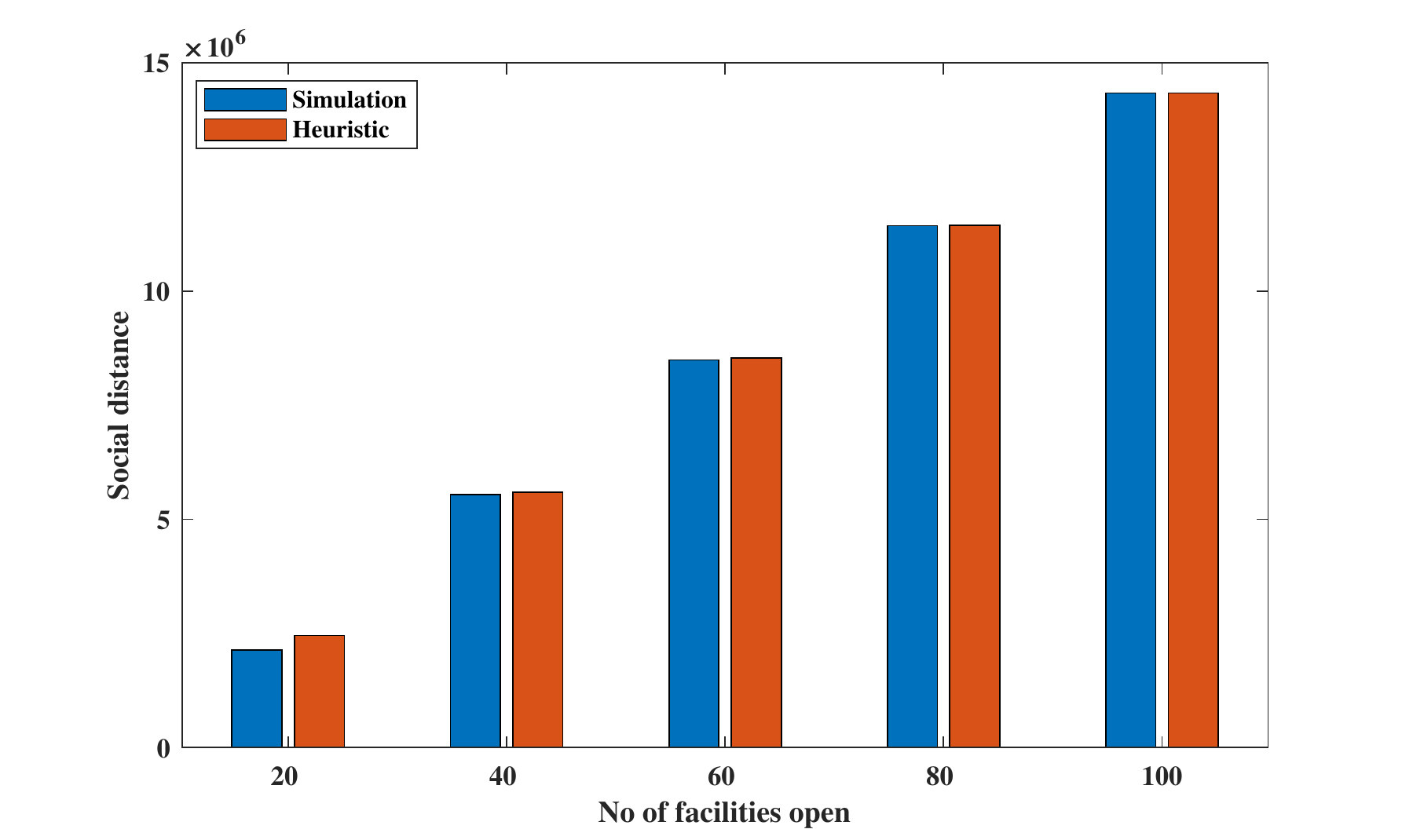} \\
(a) Grid Network ($n=60 \times 60$) & (b) Grid Network ($n=100 \times 100$) \\
\end{tabular}
\caption{Impact of the number of opened facilities on Social Distancing for Complete and Grid Network for two different sizes.}
\label{Fig:Social_Distancing}
\end{figure}

\paragraph{\textbf{Impact on the Social Distancing}}  Here, we discuss a comparative study of two different solution methodologies on social distancing. Figure \ref{Fig:Social_Distancing} shows the plots for the number of opened facilities vs. social distancing for both kinds of network topologies with two different network sizes. From the figure, it has been observed that for a fixed network size when the number of opened facilities increases, social distancing also increases. As an example For a complete network with $100$ nodes, when the number of opened facilities is $20$, the value of social distancing due to the allocation of facilities by the heuristic method is $2.4 \times 10^{6}$. When the number of opened facilities are increased to $60$, the value of social distancing also increased to $7.9 \times 10^{6}$.
\par The impact of algorithms on social distancing can also be observed from Figure \ref{Fig:Social_Distancing}. We can conclude that when the number of opened facilities is more, it does not matter which algorithm is used to locate facilities the value of social distancing will not change much. However, if the number of opened facilities is much less, the location suggested by the heuristic approach leads to more value of the social distancing. As an example, in case of the $100 \times 100$ rectangular  grid network when the number of opened facilities is $20$, the value of social distancing function due to facility placement by simulation-based approach and heuristic approach is $2.08 \times 10^{6}$ and $2.29 \times 10^{6}$, respectively. Hence, the difference is approximately $210000$.

\begin{figure}[h]
\centering 
\begin{tabular}{ c c }
\includegraphics[scale=0.4]{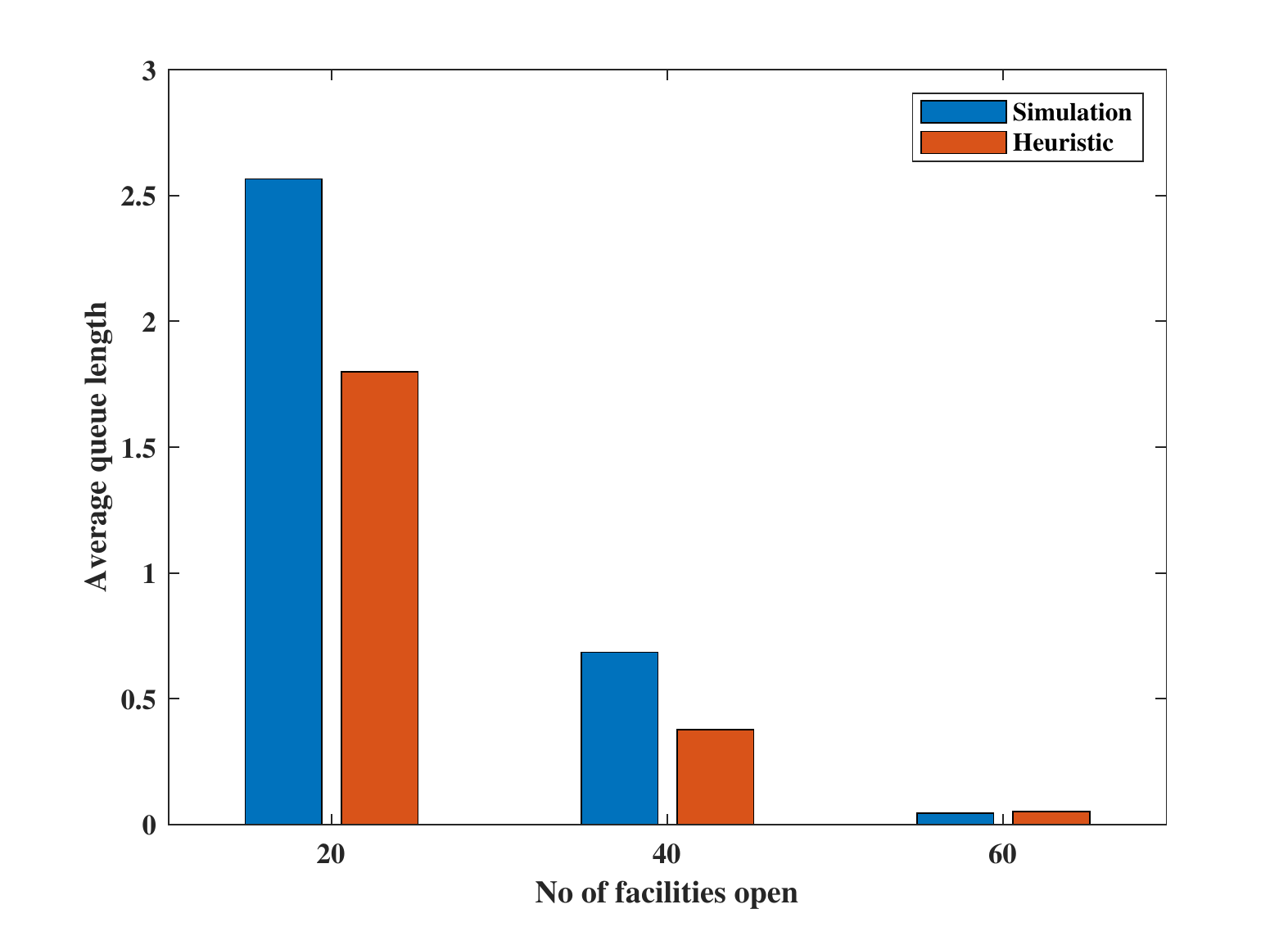} & \includegraphics[scale=0.34]{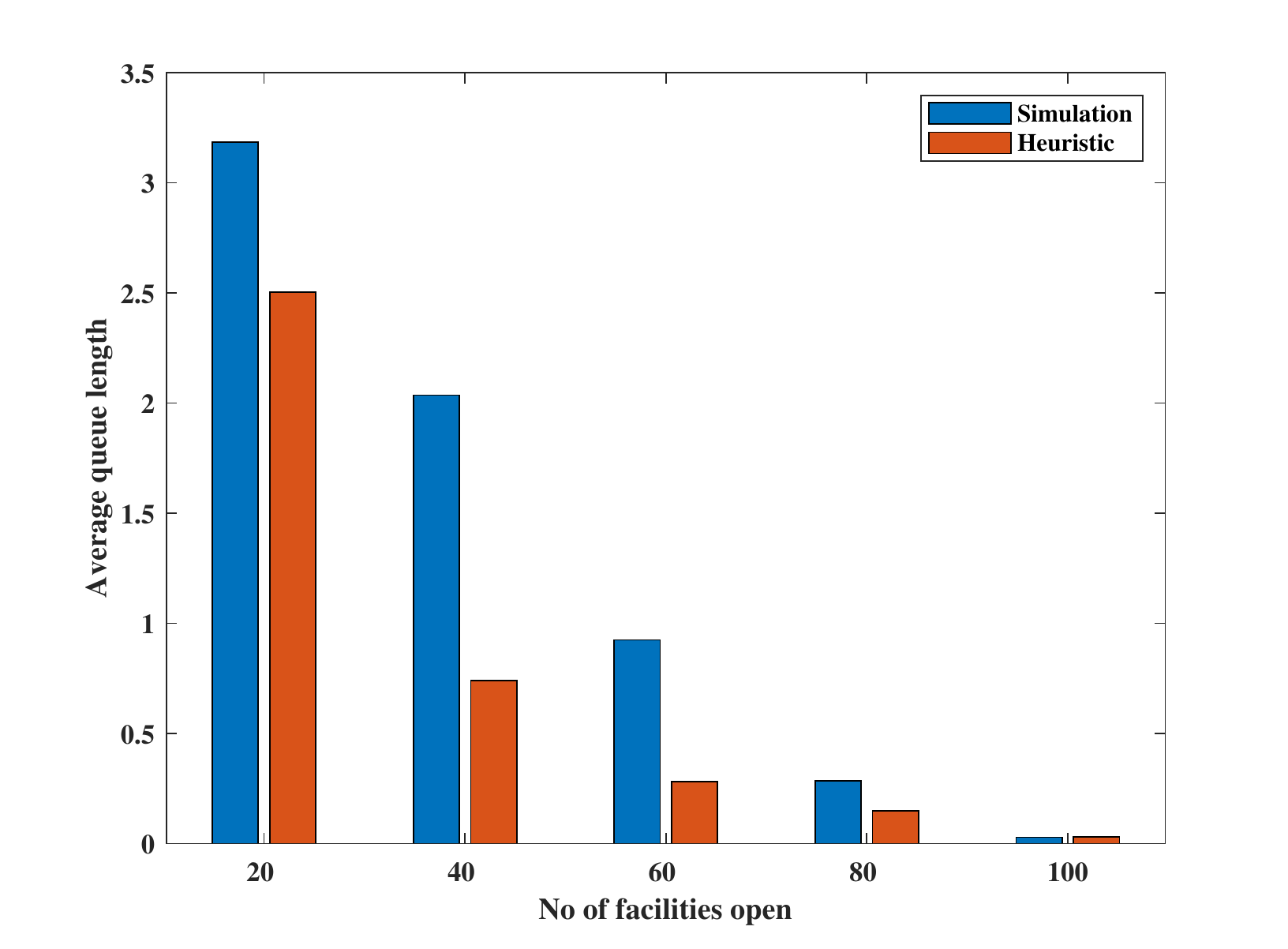} \\
(a) Complete Network ($n=60$ ) & (b) Complete Network ($n=100$)\\
\includegraphics[scale=0.4]{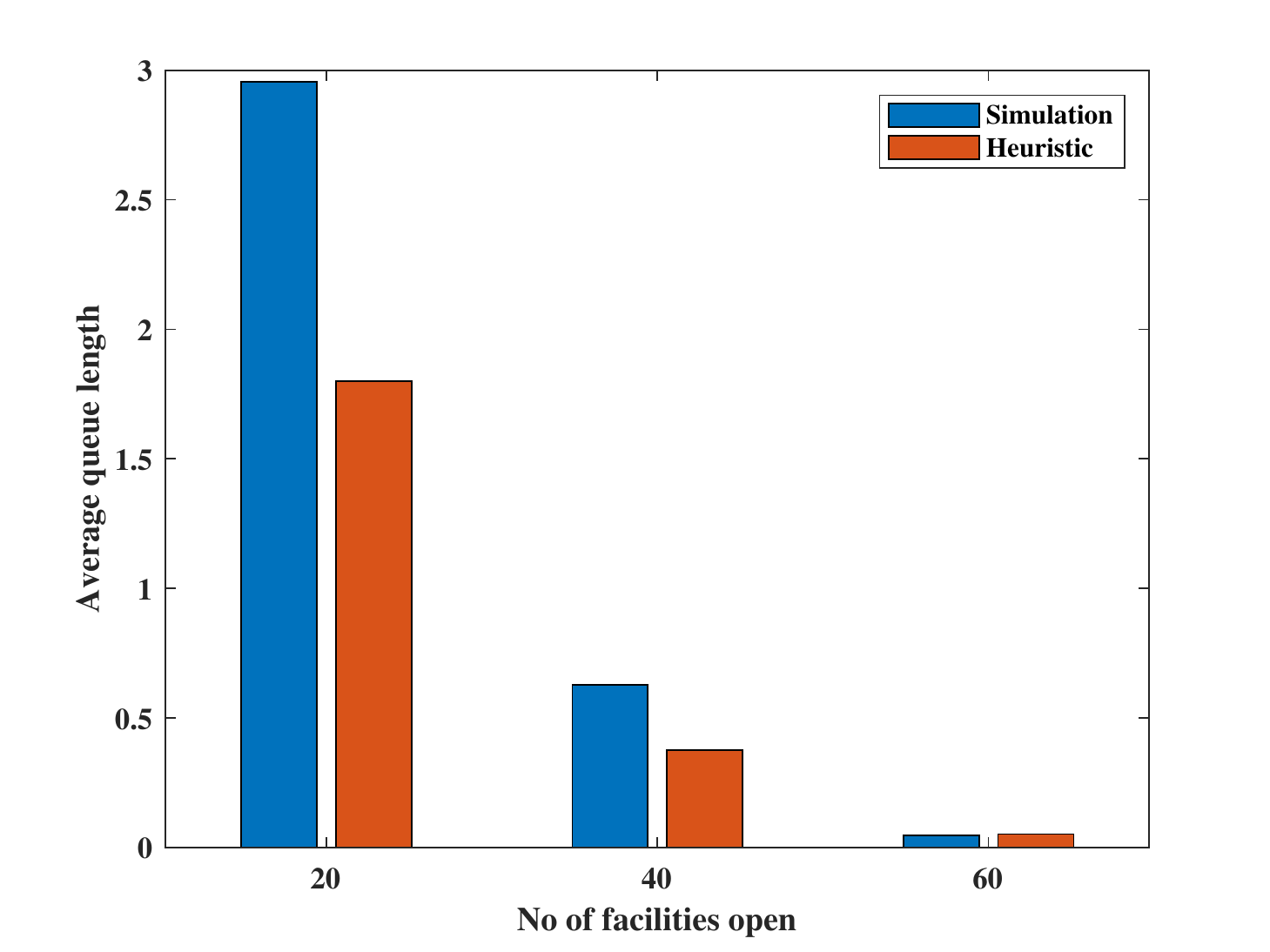} & \includegraphics[scale=0.4]{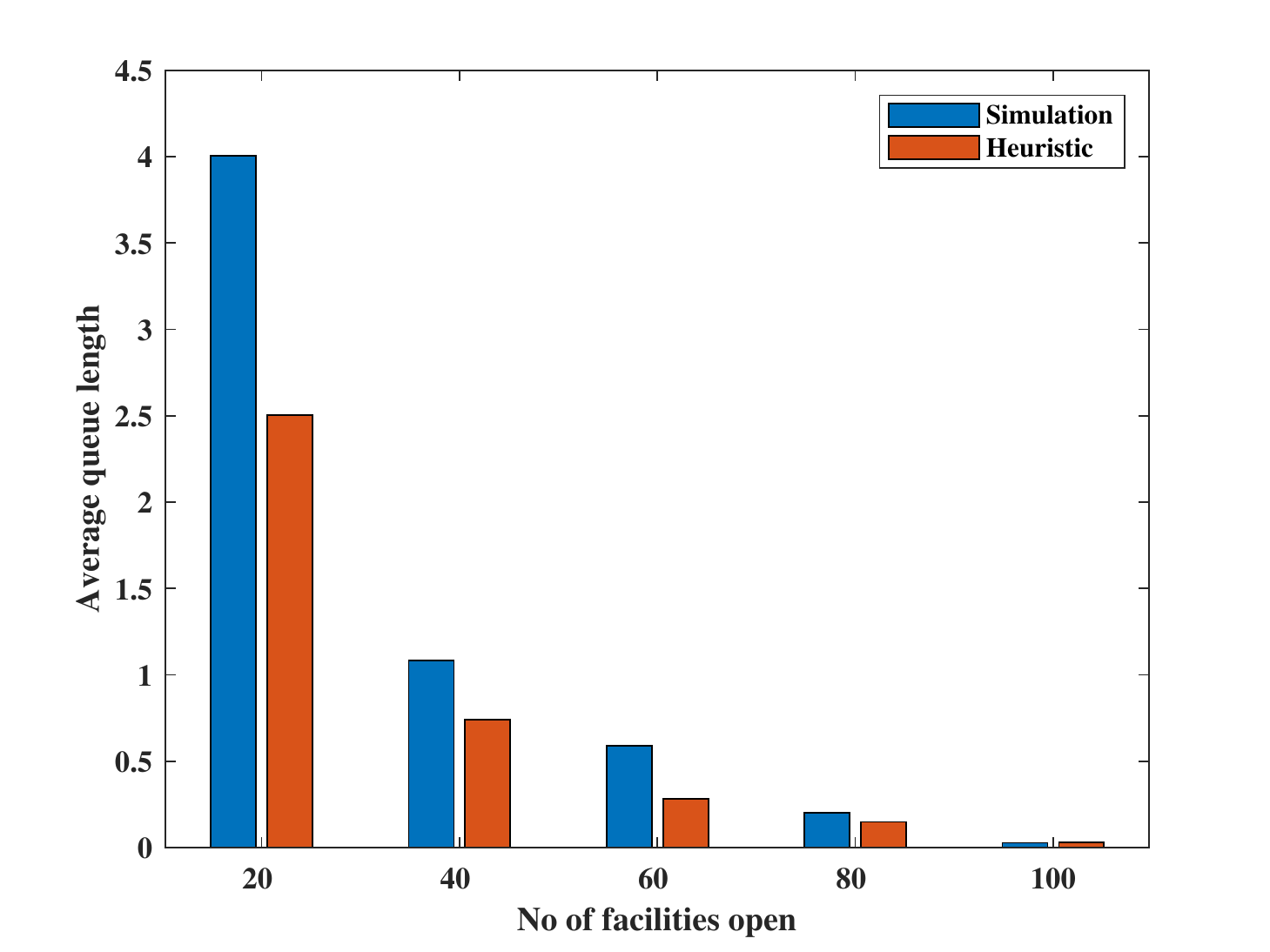} \\
(a) Grid Network ($n=60 \times 60$) & (b) Grid Network ($n=100 \time 100$) \\
\end{tabular}
\caption{Impact of the number of opened facilities on Average Queue Length for Complete and Grid Network for two different sizes.}
\label{Fig:Avg_Queue_Length}
\end{figure}

\paragraph{\textbf{Impact on Average Queue Length}} Now, we discuss the impact of the number of opened facilities on the average queue length. Figure \ref{Fig:Avg_Queue_Length} shows the number of opened facilities vs. average queue length plots for both kinds of networks for two different sizes. From the figure, it has been observed that as the number of opened facilities increases naturally, the average queue length decreases. As an example, for a complete network with $60$ nodes when the number of opened facilities is $20$, the average queue length is $2.58$ by the simulation-based approach. However, when it is increased to $40$ the average queue length drops down to $0.68$. It is important to observe that between the two proposed methodologies, the locations selected by the heuristic approach for placing the facilities lead to less average queue length. As an example, for a grid network of size $60 \times 60$ when the number of opened facilities is $20$, the the average queue length due to simulation-based and heuristic approach is $2.98$ and $1.56$, respectively.  

From the experiments, we can observe that the allocation of facilities by the heuristic approach leads to lesser queue length and consequently more value of social distancing. This is due to the fact that the heuristic approach tries to assign the people to the facilities in such a way that their loads are balanced. Due to this reason, the social distancing achieved by the heuristic solution is more.
\section{Concluding Remarks} \label{Sec:CFD}
In this paper, we have studied the problem of the Social Distancing-Based Facility Location Problem and proposed two solution methodologies. The first one is a simulation-based approach and the second one is a heuristic solution. From the experiments with synthetic datasets, we observe that the heuristic solution leads to the allocation of facilities causing more social distancing. Our study can be extended in several directions. One can consider more realistic scenarios like unknown demand, supply constraint, and many more. We can also extend our study for competitive situation where there will be more than one agency for providing the facilities of each type and their goal is to maximize their profit at the same time the social distancing needs to be maintained.

 \bibliographystyle{splncs04}
 \bibliography{Paper}

\end{document}